\documentclass[twocolumn,showpacs,showkeys,preprintnumbers]{revtex4-1}
\usepackage{amssymb}
\usepackage[paper=a4paper,dvips,top=3.7cm,left=1.2cm,right=1.2cm,]{geometry}

\usepackage{amsfonts}

\usepackage{latexsym}

\usepackage{dcolumn}

\usepackage{amsmath}

\usepackage{graphicx}

\usepackage{psfrag,pstricks}

\begin{document}

\title{Entanglement Sudden Death and Sudden Birth in Semiconductor
Microcavities}
\author{S. Abdel-Khalek$^{1,2,6}$}
\author{ Sh. Barzanjeh$^{3}$ }
\author{ H. Eleuch$^{4,5,6}$}
\affiliation{$^{1}$ Mathematics Department, Faculty of Science,
Sohag University, 82524 Sohag, Egypt} \affiliation{$^{2}$
Mathematics Department, Faculty of Science, Taif University 21974
Taif, Saudi Arabia} \affiliation{$^{3}$Department of Physics,
Faculty of Science, University of Isfahan, Hezar Jerib, Isfahan,
Iran} \affiliation{$^{4}$Institute for Quantum Science and
Engineering and Department of Physics and Astronomy, Texas A\&M
University, USA.} \affiliation{$^{5}$Physics Department, King Saud
University, P.O. Box 2455, Riyadh, Saudi Arabia}
\affiliation{$^{6}$The Abdus Salam International Centre of
Theoretical Physics, ICTP, Trieste, Italy.}
\date{\today }

\begin{abstract}

We explore the dynamics of the entanglement in a semiconductor
cavity QED containing a quantum well. We show the presence of sudden
birth and sudden death for some particular sets of the system
parameters.

\end{abstract}

 \maketitle

\section{Introduction}\label{Introduction}
Entanglement as the central feature of quantum mechanics
distinguishes a quantum system from its classical counterpart. As an
important physical resource, it has many applications in quantum
information theory. Among the well known applications of
entanglement are
superdense coding \cite{abeye}, quantum state teleportation\cite%
{benn,agrawal}. Efforts to quantify this resource are often termed
entanglement theory\cite{plenio}. Quantum entanglement also has many
different applications in the emerging technologies of quantum computing and
quantum cryptography \cite{zuko,nnrr}, and has been used to realize quantum
teleportation experimentally\cite{xian}.\newline
Quantum entanglement has attracted a lot of attention in recent years in
various kinds of quantum optical systems~\cite%
{yonac,Lajos,chan,barrada1,barrada2,barrada3}.

Several methods to quantify entanglement have been proposed. For pure
states, the partial entropy of the density matrix can provide a good measure
of entanglement. Information entropies are also used to quantify the
entanglement in quantum information \cite{shann}. In this regard the von
Neumann Entropy (NE) \cite{neum}, Linear Entropy (LE) and Shannon
information Entropy (SE) have been frequently used in treating entanglement
in the quantum systems. It is worth mentioning that the SE involves only the
diagonal elements of the density matrix and in some cases gives information
similar to that obtained from the NE and LE. On the other hand, there is an
additional entropy, namely, the Field Wehrl Entropy (FWE) \cite{wehrl}. This
measure has been successfully applied in description of different properties
of the quantum optical fields such as phase-space uncertainty \cite%
{abotalb09,mira1}, decoherence \cite{{orl},deco} etc.

The FWE is more sensitive in distinguishing states than the NE since FWE is
a state dependent \cite{mira3}. The concept of the Wehrl Phase Distribution
(WPD) has been developed and shown that it serves as a measure of both noise
(phase-space uncertainty) and phase randomization \cite{mira3}. Furthermore,
the FWE has been applied to the dynamical systems. In this respect the time
evolution of the FWE for the Kerr-like medium has been discussed in \cite%
{jex}. For the Jaynes-Cumming model the FWE gives an information on the
splitting of the Q-function in the course of the collapse region of the
atomic inversion as well as on the atomic inversion itself \cite%
{orl,obad,abotalb07}.\newline
In the current contribution we study the evolution behavior of entanglements
in a semiconductor cavity QED containing a quantum well coupled to the
environment by the FWE, generalized concurrence vector and WPD. We also
explore the situation in which the entanglement decays to zero abruptly.
Recently, Yu and Eberly \cite{yu1,yu12,yu2,yu3,yu4} showed that entanglement
loss occurs in a finite time under the action of pure vacuum noise in a
bipartite state of qubits. They found that, even though it takes infinite
time to complete decoherence locally, the global entanglement may be lost in
finite time. This phenomenon of sudden loss of entanglement has been named
as \textquotedblright entanglement sudden death\textquotedblright\ (ESD). .
Opposite to the currently extensively discussed ESD, Entanglement Sudden
Birth (ESB)\cite{yona,lop} is the creation of entanglement where the
initially unentangled qubits can be entangled after a finite evolution time.
These phenomena have recently received a lot of attention in cavity-QED and
spin chain\cite{shan,man}, and have been observed Exprimentally\cite%
{dav1,dav2}.\newline
The paper is organized as follows; Section 2 displays the physical system
and its model Hamiltonian. Section 3 devotes to evolution equations of the
system by the quantum trajectory approaching. Section 4 discusses the
entanglement due to Wehrl entropy, generalized concurrence and the Wehrl
phase distribution and section 5 supplies a conclusion and outlooks.

\bigskip

\bigskip

\section{Model}

The considered system is a quantum well confined in a semiconductor
microcavity. The semiconductor microcavity is made of a set of Bragg mirrors
with specific separation taken to be of the order of the wavelength $\lambda
$. In the system under consideration, we restricted our discussion to the
interaction of electromagnetic field with two bands in the weak pumping
regime. The electromagnetic field can make an electron transition from
valance to conduction band. This transition simultaneously creates a single
hole in the valance band which leads to generation of exciton in the system.
One can use an effective Hamiltonian without spin effects for describing the
exciton-photon coupling in the cavity as~\cite{28,29,30,31,32,33,34,35}:
\begin{eqnarray}
H &=&\hbar \omega_{p} a^{\dag}a+\hbar \omega_{e} b^{\dag}b+\imath\hbar
g^{\prime}(a^{\dag}b-b^{\dag}a)  \notag \\
&+&\hbar\alpha^{\prime} b^{\dag}b^{\dag}b b
+\imath\hbar(\varepsilon^{\prime}e^{\imath\omega t}a^{\dag}-h.c)+H_{r},
\label{1}
\end{eqnarray}
where $\omega _{p}$ and $\omega _{e}$ are the frequencies of the photonic
and excitonic modes of the cavity respectively. The bosonic operators
\textit{a} and \textit{b} are respectively describing the photonic and
excitonic annihilation operators and verifying $[a,a^{\dag }]=1;[b,b^{\dag
}]=1$. The first two terms of the Hamiltonian describe respectively the
energies of photon and exciton. The third term corresponds to the
photon-exciton coupling with a constant of coupling $g^{\prime }$. The forth
term describes the nonlinear exciton-exciton scattering due to coulomb
interaction. Where~$\alpha ^{\prime }$ is the strength of the interaction
between excitons~\cite{36,37}. The fifth term represents the interaction of
external driving laser field with the cavity, with $\varepsilon ^{\prime }$
and $\omega $ being respectively the amplitude and frequency of the driving
field. Finally, the last term describes the relaxation part of the main
exciton and photon modes. We restrict our work to the resonant case where
the pumping laser, the cavity and the exciton are in resonance ~($\omega
=\omega _{p}=\omega _{e}$). We have neglected also the photon-exciton
saturations effects in Eq.(\ref{1}). It is shown that these effects give
rise to small corrections as compared to the nonlinear exciton-exciton
scattering~\cite{29,38,39}. Furthermore, we assume that the thermal
reservoir is at the ~$T=0$ and we neglect the nonlinear dissipations~\cite%
{39b}, then the master equation can be written as ~\cite{40,41,42,42b,42c}
\begin{eqnarray}
\frac{\partial \rho }{\partial t} &=&-\imath \alpha \lbrack b^{\dag }b^{\dag
}bb,\rho ] +g[(a^{\dag }b-b^{\dag }a),\rho ]  \notag \\
&+&\varepsilon \lbrack (a^{\dag }-a),\rho ]+L\rho ,  \label{2}
\end{eqnarray}%
where $t$ is a dimensionless time normalized to the round trip time $\tau
_{c}$ in the cavity, and we normalize all constant parameters of the system
to $1/\tau _{c}$ as:~$g=g^{\prime }\tau _{c},\varepsilon =\varepsilon
^{\prime }\tau _{c},\alpha =\alpha ^{\prime }\tau _{c}$. $L\rho $ represents
the dissipation term associated with $H_{r}$ and it describes the
dissipation due to the excitonic spontaneous emission rate $\gamma /2$ and
to the cavity dissipation rate $\kappa $:
\begin{eqnarray}
L\rho &=&\kappa (2a\rho a^{\dag }-a^{\dag }a\rho -\rho a^{\dag }a)  \notag \\
&+&\gamma /2(2b\rho b^{\dag }-b^{\dag }b\rho -\rho b^{\dag }b).  \label{3}
\end{eqnarray}

\section{Evolution equations}

In the weak excitation regime $\frac{\varepsilon }{\kappa }\ll 1$, we can
neglect the non-diagonal terms $2a\rho a^{\dag }$ and $2b\rho b^{\dag }$ in
the master equation(\ref{3})~\cite{43,44}. The density matrix can then be
factorized as a pure state~\cite{37},\cite{43}-\cite{46}. We then obtain,
the following compact and practical master equation:

\begin{equation}  \label{4}
\begin{array}{rcl}
\frac{d\rho}{d t}=\frac{1}{\imath\hbar}(H_{eff}\rho-(H_{eff}\rho)^{\dag}), &
&
\end{array}%
\end{equation}
where the effective non-Hermitian Hamiltonian~$H_{eff}$ defined as

\begin{eqnarray}
H_{eff} &=&\imath \hbar g(a^{\dag }b-b^{\dag }a)+\hbar \alpha b^{\dag
}b^{\dag }bb  \notag \\
&+&\imath \hbar \varepsilon (a^{\dag }-a)-\imath \hbar \kappa a^{\dag
}a-\imath \hbar \frac{\gamma }{2}b^{\dag }b.  \label{5}
\end{eqnarray}
in which the time dependent density matrix $\rho =|\psi (t)\rangle \langle
\psi (t)|$ is a possible solution of equation(\ref{4}). Also~$|\psi
(t)\rangle $ satisfies the following equation:
\begin{equation}
\begin{array}{rcl}
\imath \hbar \frac{d|\psi (t)\rangle }{dt}=H_{eff}|\psi (t)\rangle . &  &
\end{array}
\label{6}
\end{equation}%
The essential effect of the pump field is to increase the excitation quanta
number in the cavity which allows us to neglect the term $\hbar \varepsilon
a $ in the expression of the effective non-Hermitian Hamiltonian Eq.(\ref{5}%
)~\cite{43,44,45}.\newline
We can expand~$|\psi (t)\rangle $ into a superposition of tensor product of
pure excitonic and photonic states~\cite{37},\cite{43,44,45}:
\begin{eqnarray}  \label{psi}
|\psi \left( t\right) \rangle &=&A_{00}|00\rangle +A_{10}|10\rangle
+A_{01}|01\rangle +A_{11}|11\rangle  \notag \\
&&+A_{20}|20\rangle +A_{02}|02\rangle +A_{30}|30\rangle +A_{03}|03\rangle
\notag \\
&&+A_{21}|21\rangle +A_{12}|12\rangle,
\end{eqnarray}
where ~$|ij\rangle =|i\rangle \bigotimes |j\rangle $,~is the state with $i$
photons and $j$ excitons in the cavity. We then obtain the following
differential equations for the amplitudes $A_{ij}(t)$

\begin{equation}
\left.
\begin{array}{l}
\dfrac{dA_{00}}{dt}=-\varepsilon A_{10}, \\
\dfrac{dA_{01}}{dt}=-\varepsilon A_{11}-gA_{10}-\dfrac{\gamma }{2}A_{01}, \\
\dfrac{dA_{10}}{dt}=\varepsilon \left( A_{00}-\sqrt{2}A_{20}\right)
+gA_{01}-kA_{10}, \\
\dfrac{dA_{11}}{dt}=\sqrt{2}g\left( A_{02}-A_{20}\right) -\left( k+\dfrac{%
\gamma }{2}\right) A_{11}+\varepsilon A_{01}-\varepsilon \sqrt{2}A_{21}, \\
\dfrac{dA_{20}}{dt}=\sqrt{2}gA_{11}-2kA_{20}+\sqrt{2}\varepsilon A_{10}-%
\sqrt{3}\varepsilon A_{30} , \\
\dfrac{dA_{02}}{dt}=-\sqrt{2}gA_{11}-2i\alpha A_{02}-\gamma
A_{02}-\varepsilon A_{12} , \\
\dfrac{dA_{03}}{dt}=-g\sqrt{3}A_{12}-(\dfrac{3\gamma }{2}+6i\alpha
)A_{03}(t), \\
\dfrac{dA_{30}}{dt}=\varepsilon \sqrt{3}A_{20}+g\sqrt{3}A_{21}-3kA_{30}, \\
\dfrac{dA_{12}}{dt}=\varepsilon A_{02}+g\left( \sqrt{3}A_{03}-2A_{21}\right)
-\left( k+\gamma +2i\alpha \right) A_{12}, \\
\dfrac{dA_{21}}{dt}=\varepsilon \sqrt{2}A_{11}+g\left( 2A_{12}-\sqrt{3}%
A_{30}\right) -\left( 2k+\dfrac{\gamma }{2}\right) A_{21},%
\end{array}%
\right.
\end{equation}

\bigskip

We assume that, at time~$t=0$ the vector state $|\psi (t)\rangle $ is in
vacuum state, $|\psi (t=0)\rangle =|00\rangle $:
\begin{equation}
\begin{array}{rcl}
A_{ij}(t=0)=0. &  &
\end{array}
\label{12}
\end{equation}
For pure state, the density operator can be written in term of the
wavefunction ~$|\psi (t)\rangle $ as ~$\rho _{ph,exc}=|\psi (t)\rangle
\langle \psi (t)|$. The reduced density matrices of photon-exciton system
can be written as

\begin{equation}
\begin{array}{rcl}
\rho _{ph}=tr_{exc}(|\psi (t)\rangle \langle \psi (t)|),\rho
_{exc}=tr_{ph}(|\psi (t)\rangle \langle \psi (t)|). &  &
\end{array}
\label{20}
\end{equation}%
the above equation will be used in the next sections extensively to
calculate the FWE, concurrence and WPD.

\section{Entaglement dynamics of three excitations regime}

\subsection{Wehrl entropy}

In this section, we investigate the field Wehrl entropy for the system under
consideration. Actually, the Wehrl entropy is better than the Shannon
entropy and von Neumann entropy for certain states. More illustratively the
Shannon entropy $S_{H}$ depends on the diagonal elements so that it does
not\ contain any information about the phase and can be expressed as%
\begin{equation}
S_{H}(t)=-\overset{\infty }{\underset{n=0}{\sum }}p(n,t)\ln (p(n,t)),\text{
\ }
\end{equation}%
where $p(n,t)=\left\langle n\left\vert \hat{\rho}\left( t\right) \right\vert
n\right\rangle $ is the photon number distribution. On the other hand, the
von Neumann entropy defined as $S_{N}(t)=-$Tr$(\rho (t)\ln (\rho (t)),$ can
not be used in the mixed state case.

To study of the Wehrl entropy of the photons in the case of three
excitations regime one need to calculate the Husimi $Q_{F}$ function. Which
is defined in terms of the diagonal elements of the density operator in the
coherent state basis as
\begin{equation}
Q_{ph}\left( \beta ,\Theta ,t\right) =\frac{1}{\pi }\left\vert \left\langle
\beta |\psi \left( t\right) \right\rangle \right\vert ^{2}
\label{husimientropy}
\end{equation}%
where the coherent state representation $\left\vert \beta \right\rangle =%
\overset{\infty }{\underset{n=0}{\sum }}b_{n}(\beta )\left\vert
n\right\rangle $ \ while the amplitude
\begin{equation}
b_{n}(\beta )=\frac{\exp (-\frac{\left\vert \beta \right\vert ^{2}}{2})}{%
\sqrt{n!}}\beta ^{n}\text{, \ \ \ \ \ \ }\beta =\left\vert \beta \right\vert
e^{i\Theta }
\end{equation}%
In order to compute the Husimi \textit{Q} function of the photons we
substitute the sate vector in the case of three photon excitation is given
by Eq.(\ref{psi}) into Eq.(\ref{husimientropy}) which reads

\begin{eqnarray}
&&\left.
\begin{array}{l}
Q_{ph}(\beta ,\Theta ,t)=\left\vert \overline{q_{0}(\beta )}A_{00}+\overline{%
q_{1}(\beta )}A_{10}+\overline{q_{2}(\beta )}A_{20}+\overline{q_{3}(\beta )}%
A_{30}\right\vert ^{2} \\
\ \ \ \ \ \ \ \ \ \ \ \ \ \ \ \ \ \ +\left\vert \overline{q_{0}(\beta )}%
A_{01}+\overline{q_{1}(\beta )}A_{11}+\overline{q_{2}(\beta )}%
A_{21}\right\vert ^{2} \\
\ \ \ \ \ \ \ \ \ \ \ \ \ \ \ \ \ \ +\left\vert \overline{q_{0}(\beta )}%
A_{02}+\overline{q_{1}(\beta )}A_{12}\right\vert ^{2}+\left\vert \overline{%
q_{0}(\beta )}A_{03}\right\vert ^{2}%
\end{array}%
\right.   \notag \\
&&
\end{eqnarray}

Now, we may calculate the Wehrl entropy. The concept of the classical-like
Wehrl entropy (FWE) is a very informative measure describing the time
evolution of a quantum system. The Wehrl entropy, introduced as a classical
entropy of a quantum state, can give additional insights into the dynamics
of the system, as compared to other entropies. The Wehrl classical
information entropy is defined as~\cite{wehrl}
\begin{equation}
S_{W}\left( t\right) =-\int_{0}^{2\pi }\int_{0}^{\infty }Q_{ph}\left( \beta
,\Theta ,t\right) \ln Q_{ph}\left( \beta ,\Theta ,t\right) \left\vert \beta
\right\vert d\left\vert \beta \right\vert d\Theta
\end{equation}%
We point out that the state vector coefficients and \textit{Q}-function both
are normalized at time steps as follows
\begin{eqnarray}
\sum \left\vert A_{ij}\right\vert ^{2} &=&1\ \ \  \\
\int_{0}^{2\pi }\int_{0}^{\infty }Q_{ph}\left( \beta ,\Theta ,t\right)
\left\vert \beta \right\vert d\left\vert \beta \right\vert d\Theta  &=&1
\end{eqnarray}%
\begin{figure}[tbp]
\centering\includegraphics[width=3.5in]{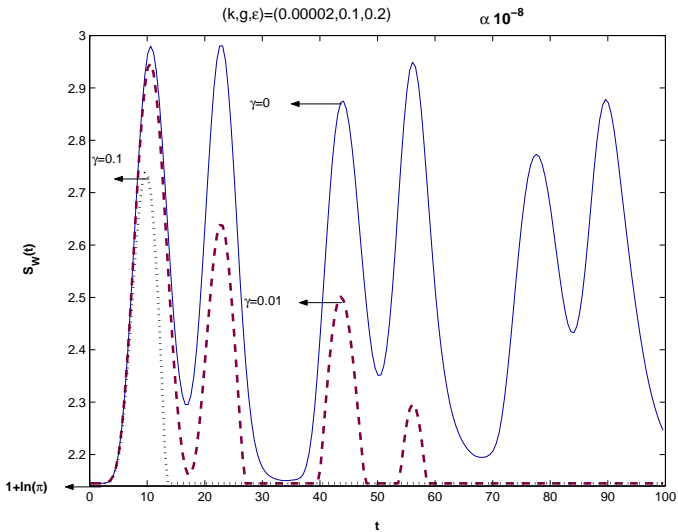}
\caption{The time evolution of the Wehrl entropy $S_{W}(t)$ \ for $\protect%
\alpha =10^{-8}$, ($k,g,\protect\varepsilon $)=($0.00002,0.1,0.2$)
and with different values of \ $\protect\gamma $.}
\end{figure}

\begin{figure}[tbp]
\centering\includegraphics[width=3.5in]{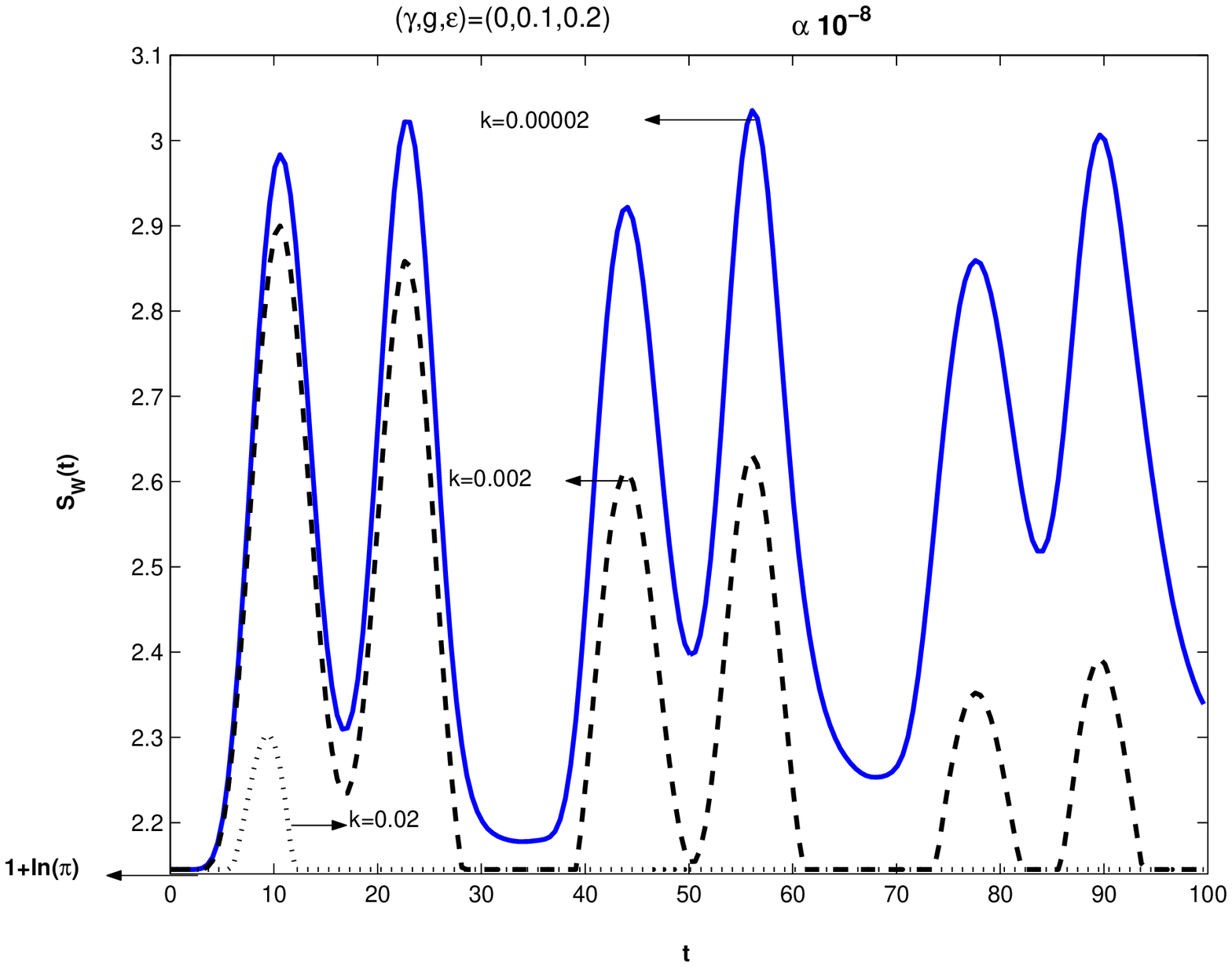}
\caption{The time evolution of the Wehrl entropy $S_{W}(t)$ \ for $\protect%
\alpha =10^{-8}$, ($\protect\gamma ,g,\protect\varepsilon
$)=($0,0.1,0.2$) and with different values of \ $k.$}
\end{figure}
\begin{figure}[tbp]
\centering\includegraphics[width=3.5in]{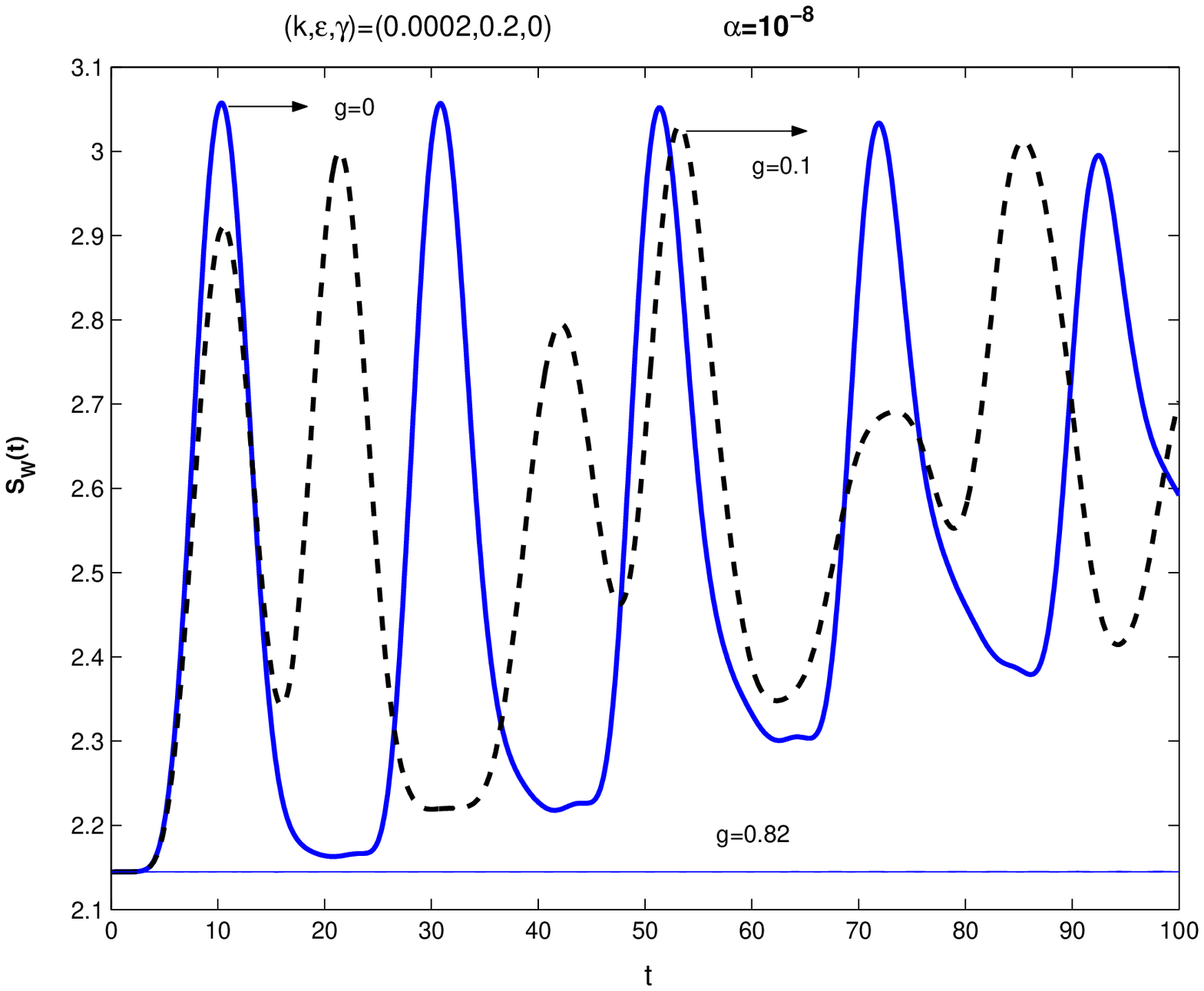}
\caption{The time evolution of the Wehrl entropy $S_{W}(t)$ \ for $\protect%
\alpha =10^{-8}$, ($k,\ \protect\gamma ,\protect\varepsilon $%
)=(0.0002,0,0.2) and with different values of $g$.}
\end{figure}
\begin{figure}[tbp]
\centering\includegraphics[width=3.5in]{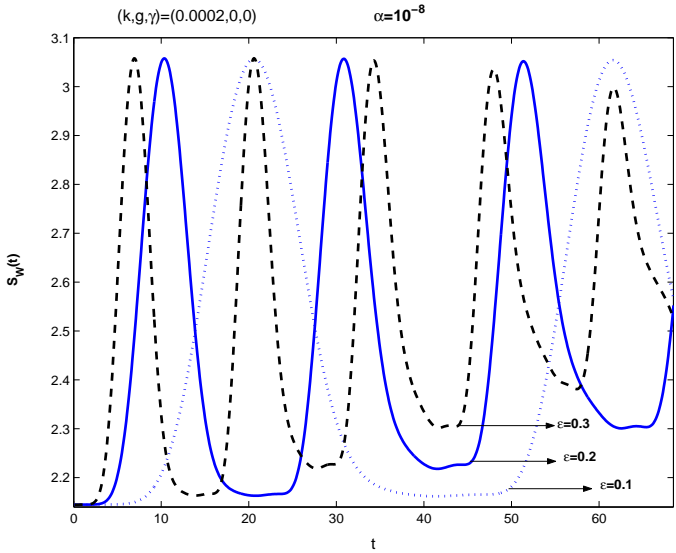}
\caption{The time evolution of the Wehrl entropy $S_{W}(t)$ \ for $\protect%
\alpha =10^{-8}$, ($g,\protect\gamma ,k$)=($0,0,0.0002$) and with
different values of $\protect\varepsilon $.}
\end{figure}
To explore the influence of decoherence on the dynamical behavior of the
Wehrl entropy, we have plotted the time evolution of the photon Wehrl
entropy~$S_{W}(t)$ as a function of time~$t$ for different values of the
coupling constant $\gamma $ and the cavity dissipation rate $k$ in
Figures.(1) and (2)\newline
Figure.(1) shows the influences of excitonic spontaneous emission rate~$%
\gamma $ on the Wehrl entropy (WE). By increasing $\gamma $ the (WE)
decreases. Furthermore, similar effect for the cavity dissipation rate $k$
can be observed in the Fig.(2). It is worth to note that for large values of
$k$ the (WE) decreases abruptly much faster than Fig.(1). The increasing of $%
k$ or $\gamma $ enhances the decoherence in the system and consequently
causes the destruction of entanglement in the system. To have further
insight, we plot in Fig.(3) and Fig.(4) the Wehrl entropy for different
values of the coupling constant ~$g$ and the amplitude of the driving field $%
\epsilon $, respectively. By increasing the coupling constant $g$ the
frequency oscillation of the Wehrl entropy increases . This effect is also
observed in the autocorrelation function ~\cite{eleuch} and in two photon
excitations \cite{barzanjeh}.

\subsection{ Generalized concurrence}
\begin{figure}[tbp]
\centering\includegraphics[width=3.5in]{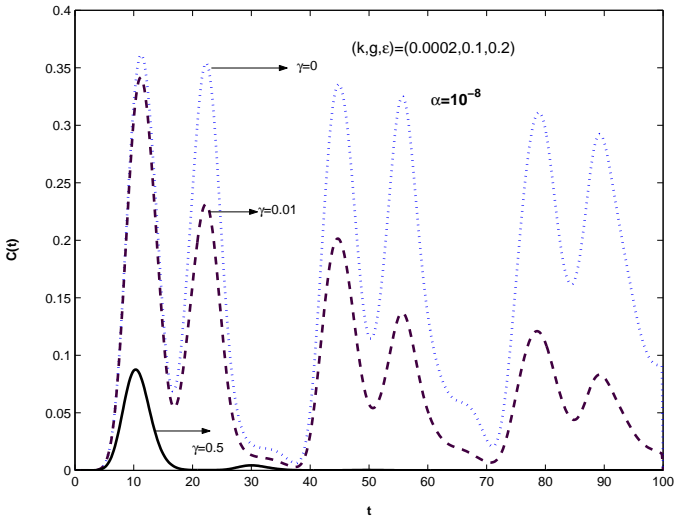}
\caption{The time evolution of the Concurrence $C(t)$ \ for $\protect\alpha %
=10^{-8}$, ($k,g,\protect\varepsilon $)=(0.0002,0.1,0.2) and with
different values of \ $\protect\gamma $.}
\end{figure}
\begin{figure}[tbp]
\centering\includegraphics[width=3.5in]{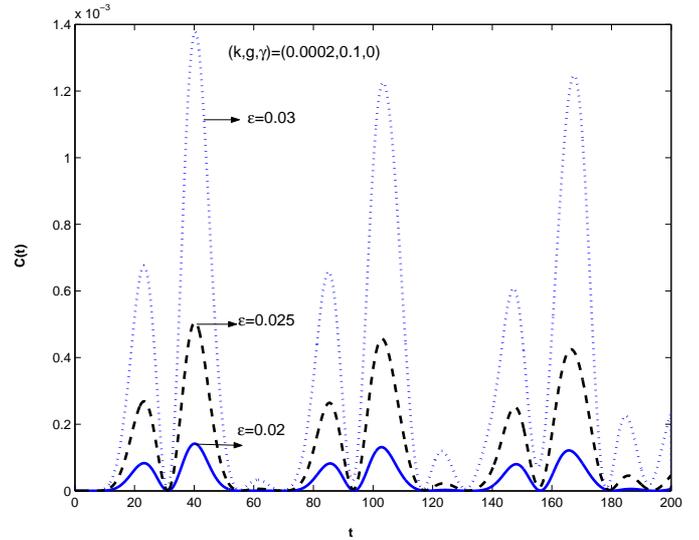}
\caption{The time evolution of the Concurrence $C(t)$ \ for $\protect\alpha %
=10^{-8}$, ($k,g,\protect\gamma $)=(0.0002,0.1,0) and with different
values of \ $\protect\varepsilon $.}
\end{figure}
\begin{figure}[tbp]
\centering\includegraphics[width=3.5in]{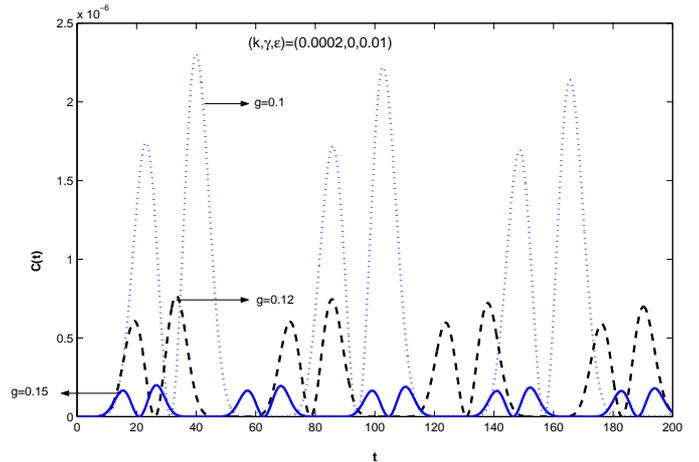}
\caption{The time evolution of the Concurrence $C(t)$ \ for $\protect\alpha %
=10^{-8}$, ($k,\ \protect\gamma ,\protect\varepsilon
$)=(0.0002,0,0.01) and with different values of $g$.}
\end{figure}
To study of entanglement for pure states usually the partial entropy of the
density matrix is a good measure of entanglement which reads
\begin{equation}
\begin{array}{rcl}
E(\psi )=-tr(\rho ~Ln\rho )=-\sum\limits_{i}{(\lambda _{i}~Ln\lambda _{i})},
&  &
\end{array}
\label{entropy}
\end{equation}%
where~$\rho $ is the reduced density matrix,~$\lambda _{i}$ is the \textit{i}
th eigenvalue of~$\rho $. In the case of a two qubit mixed state~$\rho $,
the concurrence of Wootters can be used as a measure of entanglement which
is given by\cite{48}

\begin{equation}
\begin{array}{rcl}
C(\rho )=max(0,\lambda _{1}-\lambda _{2}-\lambda _{3}-\lambda _{4}), &  &
\end{array}
\label{concurrence}
\end{equation}%
in which the~$\lambda _{i}$ are the square roots of eigenvalues in
decreasing order of ~$\sqrt{\rho }\tilde{\rho}\sqrt{\rho }$ with ~$\tilde{%
\rho}=(\sigma _{y}\bigotimes \sigma _{y})\rho ^{\ast }(\sigma _{y}\bigotimes
\sigma _{y})$. Recently, some extensions have proposed for definition of
concurrence in the case of an arbitrary bipartite pure state~$|\psi \rangle
=\sum\limits_{i=1}^{N_{1}}{\sum\limits_{j=2}^{N_{2}}{a_{ij}\left\vert {%
e_{i}\otimes e_{j}}\right\rangle }}$ as~\cite{49,50}

\begin{equation}  \label{conc}
\begin{array}{rcl}
C(\psi ) = \sqrt{\frac{2N}{N-1}}\sqrt {\sum\limits_{i < j}^{N1} {%
\sum\limits_{k < l}^{N2} {\left| {a_{ik} a_{jl} - a_{il} a_{jk} } \right|^2 }
} }. &  &
\end{array}%
\end{equation}
where $N=min(N_1,N_2)$.\newline
Here, we deal with a pure state~$|\psi\rangle\in\mathbb{C}^4\bigotimes%
\mathbb{C}^4$ so that, $N=N_1=N_2=4$. To study the time evolution of the
concurrence in the case of three excitations we substitute the state vector (%
\ref{psi}) into the Eq.(\ref{conc}), thus we obtain
\begin{equation}
C(\psi )=\sqrt{\frac{8}{3}\Upsilon (t)}
\end{equation}

where $\Upsilon (t)$ is given by
\begin{eqnarray}
&&\left.
\begin{array}{r}
\Upsilon (t)=\left\vert A_{00}A_{21}-A_{01}A_{20}\right\vert ^{2}+\left\vert
A_{03}A_{10}\right\vert ^{2}+\left\vert A_{11}A_{03}\right\vert ^{2} \\
+\left\vert A_{00}A_{11}-A_{01}A_{10}\right\vert ^{2}+\left\vert
A_{03}A_{12}\right\vert ^{2}+\left\vert A_{02}A_{20}\right\vert ^{2} \\
+\left\vert A_{01}A_{12}-A_{11}A_{02}\right\vert ^{2}+\left\vert
A_{20}A_{03}\right\vert ^{2}+\left\vert A_{02}A_{21}\right\vert ^{2} \\
+\left\vert A_{00}A_{12}-A_{12}A_{10}\right\vert ^{2}+\left\vert
A_{03}A_{21}\right\vert ^{2}+\left\vert A_{01}A_{30}\right\vert ^{2} \\
+\left\vert A_{10}A_{21}-A_{11}A_{20}\right\vert ^{2}+\left\vert
A_{02}A_{30}\right\vert ^{2}+\left\vert A_{03}A_{30}\right\vert ^{2} \\
+\left\vert A_{12}A_{20}\right\vert ^{2}+\left\vert A_{21}A_{12}\right\vert
^{2}+\left\vert A_{11}A_{30}\right\vert ^{2} \\
+\left\vert A_{12}A_{30}\right\vert ^{2}+\left\vert A_{21}A_{30}\right\vert
^{2}%
\end{array}%
\right.  \notag \\
&&
\end{eqnarray}%
\textbf{\ } We plot the time dependent concurrence vector as a function of
time $t$ for three values of $\gamma $ in Fig.(5). As it seen any increasing
of $\gamma $ leads to decreasing of entanglement similar to the Wehrl
entropy. Furthermore, an interesting cases are observed in the Fig.(6)-(7).
These figures show that the concurrence is periodic in the domain of time.
Moreover, unlike the large values of $\varepsilon $, figure (6) shows that
entanglement can fall abruptly to zero (the two lower curves in the figure)
for small values of $\varepsilon $($\varepsilon =0.025$ and $\varepsilon
=0.02$), and remains zero for a period of time before entanglement recovers.
The abrupt disappearance of entanglement that persists for a period of time
is referred to as \textit{sudden death of entanglement} (ESD)\cite%
{yu1,yu12,newref1,newref2} and also the fast appearance of entanglement
after a while is called sudden birth of entanglement(ESB)\cite{yona,lop}.
The length of the time interval for the zero entanglement is dependent on
the values of $\varepsilon $. The smaller values of $\varepsilon $, the
longer the state will stay in the disentangled separable state. Furthermore
we show that the (ESD) and (ESB) can be affected strongly by the coupling
constant $g$. As it is seen from figure.(7) the (ESD) and (ESB) can be
enhanced by increasing of coupling constant $g$. We point out that, in the
Figures.(5)-(7) we assume the zero value of the excitonic spontaneous
emission rate ~$\gamma =0$, thus one important reason for the (ESD) is the
interaction of system with its surrounding.

\subsection{Wehrl Phase distribution}
\begin{figure}[tbp]
\centering\includegraphics[width=3.5in]{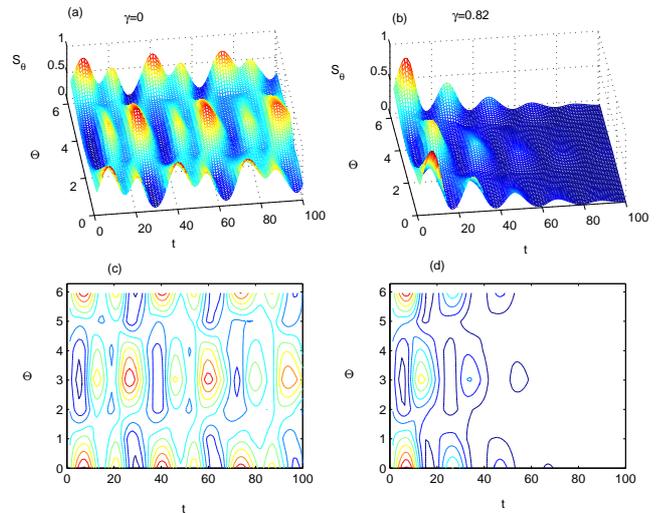}
\caption{The time evolution of the Wehrl PD $S_{\Theta }(t)$ \ for $\protect%
\alpha =10^{-8}$, ($k,g,\protect\varepsilon $)=(0.0002,0,0.2) and
with different values of \ $\protect\gamma $.}
\end{figure}
\begin{figure}[tbp]
\centering\includegraphics[width=3.5in]{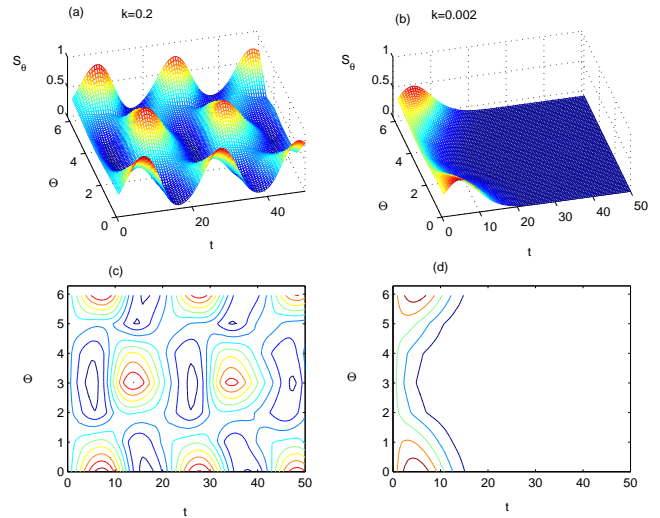}
\caption{The time evolution of the Wehrl PD $S_{\Theta }(t)$ \ for $\protect%
\alpha =10^{-8}$, ($\protect\varepsilon ,g,\protect\gamma
$)=(0.2,0,0) and with different values of \ $k$.}
\end{figure}
The Wehrl phase distribution (Wehrl PD), defined to be the phase density of
the Wehrl entropy \cite{abotalb09,mira3}, i.e.,

\begin{equation}
S_{\Theta }(t)=-\int Q_{ph}\left( \beta ,\Theta ,t\right) \ln Q_{ph}\left(
\beta ,\Theta ,t\right) \left\vert \beta \right\vert d\left\vert \beta
\right\vert ,  \label{wpd}
\end{equation}%
where $\Theta =\arg \left( \beta \right) $ and $Q_{ph}\left( \beta ,\Theta
,t\right) $ is given in Eq.(\ref{husimientropy}).

Based on Eq. (\ref{wpd}), we present some interesting results for the
effects of excitonic spontaneous emission and the dissipative rate of the
cavity on the entanglement behavior in the point of view of Wehrl PD. It is
observed that when $\gamma =0$ (see fig.8(a)) $S_{\Theta }(t) $ oscillates
between maximum and minimum peaks which is an indication of ESB and ESD. For
$\gamma\neq0 $ the situation is completely different, the excitonic
spontaneous emission destroys the entanglement (see fig.8).

Now, we would like to answer the question: How $S_{\Theta }(t)$, is
influenced by the cavity dissipation? \ For this purpose, we take two
different values of $k$ in fig.9. For small values of $k$ $S_{\Theta }(t)$
oscillates but when $k$ increases $S_{\Theta }(t)$ decreases quickly without
oscillation(see figure 9(b)). This shows a one-to-one correspondence between
the behavior of $S_{\Theta }(t)$ and the Wehrl entropy or concurrence which
opens the door for using $S_{\Theta }(t)$ as an entanglement measure.

\section{Conclusion}

In this paper we have studied the dynamical behavior of the quantum
entanglement for a semiconductor microcavity containing a quantum well. The
system is pumped with weak laser amplitude. We studied the time evolution of
entanglement between the photon-exciton by the field Wehrl entropy,
generalized concurrence and Wehrl phase distribution. Our results show that
the new features such as entanglement sudden death and entanglement sudden
birth can be reported for specific values of the cavity dissipation rate and
the excitonic spontaneous emission rate.

\end{document}